\documentclass[12pt]{article}

\usepackage{booktabs}
\usepackage{dcolumn} 
\usepackage{epstopdf}
\usepackage{graphicx}
\usepackage{hyperref}
\usepackage{longtable} 
\usepackage{natbib}
\usepackage{rotating}
\usepackage{tabularx}
\usepackage{amsmath}
\usepackage{setspace}
\usepackage{caption}

\usepackage[super]{nth}  
\hypersetup{
  colorlinks = TRUE,
  citecolor=blue,
  linkcolor=red,
  urlcolor=black
}

\hypersetup{colorlinks = TRUE, citecolor=blue, linkcolor=red, urlcolor=black}

\newcommand{\starlanguage}{Significance indicators: $p \le 0.05:*$, $p \le 0.01:**$ and $p \le .001:***$.}

\newcommand{\MU}{\hat{\mu}}  
\newcommand{\SD}{\hat{\sigma}}

\newcommand{\firstWeek}{2014-01-08}
\newcommand{\lastWeek}{2015-03-19}
\newcommand{\numCountries}{9}
\newcommand{\numObs}{549}
\newcommand{\numWeeks}{61}
\newcommand{\AppElasticity}{1.4}
\newcommand{\AppElasticityTenPct}{14}
\newcommand{\AppActiveElasticityTenPct}{13}

\newcommand{\WBElasticityTenPct}{-4.69}
\newcommand{\WBEll}{-0.666}
\newcommand{\WBErl}{-0.272}
\newcommand{\WREll}{-0.12}
\newcommand{\WRErl}{0.098}

\newcommand{\HiresElasticity}{1.4}
\newcommand{\RussianAppPoolElasticity}{0.71}

\newcommand{\PctRussianHiresPre}{1.4}
\newcommand{\PctRussianHiresPost}{2.4}

\newcommand{\PanelStartIndv}{2014-01-07}
\newcommand{\PanelEndIndv}{2015-03-17}
\newcommand{\CountryList}{Canada, Russia, United Kingdom and the United States}
\newcommand{\numObsInd}{6,733,377}

\newcommand{\numPeriods}{63}

\newcommand{\SampleStart}{2013-01-01}
\newcommand{\SampleEnd}{2014-01-01}
\newcommand{\SampleSize}{106,879}
\newcommand{\SampleSizeRussians}{5,789}

\newcommand{\HoursSScop}{$\MU=18.24$;$\SD = 20.20$}
\newcommand{\AppsSScop}{$\MU=4.63$;$\SD = 6.13$}
\newcommand{\WageSScop}{$\MU=21.62$;$\SD = 15.75$}
\newcommand{\EffectAppElasticity}{0.21}
\newcommand{\EffectActiveExtensive}{0.24}

\newcommand{\EffectHoursElasticity}{0.16}
\newcommand{\EffectWageElasticity}{-0.03}

\newcommand{\MaxHours}{100}

\newcommand{\NumRusHiredWeekFirst}{92}
\newcommand{\NumRusHiredWeekLast}{234}

\begin{document} 

\title{{The Ruble Collapse in an Online Marketplace:}\\ Some Lessons for Market Designers}


\date{\today}

\author{John J. Horton\footnote{Thanks to
    Bo Cowgill,
    Josh Angrist,
    David Autor, 
    Alan Benson, 
    Richard Zeckhauser,
    Alex Tabarrok,
    Jonathan Meer,
    Apostolos Filippas, 
    Simon J\"{a}ger,
    Paul Oyer, 
    Bob Hall,
    Christina Patterson, and Judith Chevalier for helpful comments.
    Latest draft available at \href{http://www.john-joseph-horton.com/papers/russia_shock.pdf}{http://www.john-joseph-horton.com/papers/russia\_shock.pdf
    This research was declared exempt by the MIT COUHES, Project E-2154.
    }. 
  } \\ MIT Sloan \& NBER}

\maketitle

\begin{abstract}
  \noindent The sharp devaluation of the ruble in 2014 increased the real returns to Russians from working in a global online labor marketplace, as contracts in this market are dollar-denominated.
  Russians clearly noticed the opportunity, with Russian hours-worked increasing substantially, primarily on the extensive margin---incumbent Russians already active were fairly inelastic.
  Contrary to the predictions of bargaining models, there was little to no pass-through of the ruble price changes in to wages.
 There was also no evidence of a demand-side response, with buyers not posting more ``Russian friendly'' jobs, suggesting limited cross-side externalities. 
 The key findings---a high extensive margin elasticity but low intensive margin elasticity; little pass-through into wages; and little evidence of a cross-side externality---have implications for market designers with respect to pricing and supply acquisition.
\end{abstract} 
\onehalfspacing 
\newpage

\section{Introduction}
After a marketplace business has overcome the ``cold start'' or ``chicken-and-egg'' problem \citep{rochet2006, caillaud2006}, it has to decide much to invest in acquiring more buyers or sellers.
This investment can mean subsidizing participation for one side or the other, lowering the platform's take rate, building features that are attractive to market participants, or spending on marketing.
How should a platform think about these decisions?   
Will these expenditures be effective at attracting more participants?
What will the additional participants do to prices and quantities in the marketplace---and hence platform revenue under common price structures?
How much will acquiring one side stimulate usage from the other side? 

In this paper, I report the results of a natural experiment that approximated a massive subsidy to a select group of sellers in an online labor market \citep{horton2010online}.
This ``subsidy'' was not a platform decision, but rather was caused by the collapse of the Russian ruble, which nearly doubled the real returns to Russians from working in the marketplace, as contracts in the marketplace are dollar-denominated.
This ruble collapse was not due to a general appreciation of the US dollar, but rather had Russia-specific causes, namely the imposition of sanctions following Russia's annexation of Crimea and military intervention in Ukraine; a fall in oil prices is also thought to have played a role. 
As there are few employers from Russia in the marketplace, the ruble collapse was not a demand shock.\footnote{
  I use the terms ``worker'' and ``employer'' to be consistent with the labor literature, and not as a commentary on the relationships created on the platform. 
}

To estimate effects of the ruble collapse, I primarily use a difference-in-differences design, where Russians are the treated group and non-Russians are the control group.
Of course, as the market is international, Russians and non-Russians compete, creating the possibility of spill-overs.
To address this concern, I exploit the fact that individuals from a particular country tend to specialize in the same areas of work, and so countries and workers varied in how ``exposed'' they were to the Russian influx.
As it is, the degree of competition seems to matter little, which is perhaps unsurprising given that Russians were relatively small players in the market. 

I analyze the Russian response at two levels: the ``macro'' country level and the ``micro'' individual level, using a sample of workers active pre-collapse.
At the macro level, a 10\% rise in the value of the US dollar measured in rubles led to a \AppElasticityTenPct{}\% increase in the number of applications sent per week by Russians.
The number of Russians hired increase by a similar amount. 
These macro estimates are about 5 times larger than the equivalent micro elasticity estimates.
The reason for the gap is that the macro estimates capture a substantial extensive margin response.\footnote{
  See \cite{chetty2011micro} for the macro/micro reconciliation in conventional settings.
}

On the intensive margin, the individual elasticity of hours-worked is \EffectHoursElasticity{}, though there is substantial individual heterogeneity in labor supply.
I find that the individual hours-worked and application elasticities are similar in magnitude.
This similarity implies Russian workers scale up their search intensity and hours-worked in the same proportion, with no apparent loss of job search efficiency.

Russian wages did not fall by an economically significant amount.
In the micro panel of incumbent Russians, the elasticity of the average wage to the US dollar priced in rubles is just \EffectWageElasticity{}, implying little to no pass-through.   
Although there is a slight decline in wages in the macro panel, this can be explained by changes in composition.

The paper provides a case study of how workers respond to changes in real returns on a marketplace, and offers insight how the market ``works.''
But it also has several practical implications relevant to market designers.

All marketplace designers worry about attracting buyers and sellers to their platform.
The ruble episode shows that the supply side of the market is collectively quite elastic with respect to the financial pay-off obtained on the platform.
Although it is perhaps no surprise that greater earning possibilities would influence sellers, if we thought horizontal preferences were very important in sellers choosing a marketplace \citep{hossain2011}, then effects could have been small.
They were not.
Furthermore, with labor, it also not always widely accepted that labor supply curves are even upward sloping, due to target earning \citep{camerer1997labor} or income effects.
Neither concern seems to have been borne out in practice. 
In short, money matters.

Also relevant to would-be market designers is the source of additional supply. 
Most of the supply increase was on the extensive margin, rather than the intensive margin---Russians already active on the platform only expanded their hours-worked slightly.
This has implications for where more supply can (cheaply) come from---namely from new suppliers rather than hoping existing suppliers can increase output.
For some marketplaces, we might expect a different result---particularly when sellers are firms---but when sellers in a marketplace are individual workers providing their own labor, they are inherently supply-constrained \citep{horton2019buyer}.
In this case, increases from already active sellers are likely to be expensive. 

For platforms that tax transactions, a concern with adding supply is that is might lower prices and thus potentially revenue, depending on changes in marketplace quantities.
This concern was not borne out---added supply did not discernibly lower wages, either at the market level or for the subsidized sellers.
The likely reason wages did not fall is that Russians are apparently good substitute for other workers on the platform.
As such, the increase in Russians was not a large change in supply, properly defined. 
Although the fraction of Russians increased dramatically, they are not a large component of the market as a whole, and so it was not a market-moving shock: \PctRussianHiresPre{}\% of hired workers at the start of the period I examine were Russian, compared to \PctRussianHiresPost{}\% at the end.

One might have thought Russians would have lowered their wage asks in order to obtain more work, consistent with bargaining models of price formation, but this was not the case.
The market was better characterized by the competetive market model in which workers and employers are price-takers.

The classic way to think of price-setting in a two-sided marketplace, is that the two sides---buyers and sellers---exert different cross-side externalities, which in turn should affect the price the platform wants to charge, as well as the cost of servicing that side and that side's elasticity of demand \citep{armstrong2006competition}.   
In the case of the ruble collapse, there is no evidence of a demand response---there was no discernible demand increase in ``Russian compatible'' job posts.
This might suggest cross-side externalities might be weak in an established market, but recall the lack of a price effect.
As there was no pass-through of lower wages, there was no obvious additional benefit accruing to buyers from more Russians.
With a more massive shock that moved on-platform prices, this would perhaps not be the case.\footnote{  
  Not all cross-side externalities show up in prices---\cite{hall2017labor} shows that when the number of drivers in a city increases, wait-times falls. 
}

If the market were different and Russians offered some kind of benefit to would-be employers that was distinctive and differentiated, then perhaps we would have observed some demand response. 
However, it is not clear that even for large ``shocks'' we would see large changes if the supply curve is \emph{de facto} horizontal.
In a fairly commodified market with low barriers to entry and lots of atomistic sellers, marketplace supply could be inherently highly elastic.
Although this paper only adds some evidence for this point of highly elastic supply curves, results from several other online marketplaces all point towards highly elastic supply.

In terms of related literature, there are several papers that have a plausibly exogenous increase in sellers due to policy changes or market mergers. 
For example, \cite{reshef2020} examines the role of additional sellers on a marketplace, finding that the market expansion effects were generally positive for incumbents, but not for lower quality incumbents who experience a business stealing effect.
\cite{farronato2020dog} examines the effects of a platform acquisition, finding that although some incumbent users benefited. There was enough platform differentiation that users of the acquired platform experienced worse outcomes.  
There are also papers looking at entirely new kind of competitor, as in \cite{seamans2014}, or a new kind of product offering in an ecosystem, as in \cite{li2017}.
There are no papers I am aware of that have a distinct group of sellers entering who are not necessarily strongly differentiated form incumbents.

The rest of the paper is organized as follow.
I first discuss the empirical context in Section~\ref{sec:empirical_context}.
Results at the ``macro'' country-level are presented in Section~\ref{sec:country_results}.
Results at the ``micro'' individual level are presented in Section~\ref{sec:individual_results}.
In Section~\ref{sec:outcomes_for_russian_vacancies}, I show how little effect the Russian influx had on the demand side of the market.
Section~\ref{sec:discussion} discusses the results and Section~\ref{sec:conclusion} concludes. 

\section{Empirical context} \label{sec:empirical_context}

The empirical context is a large online labor market \citep{horton2010online, horton2017digital, agrawal2015digitization}.
In these markets, firms hire workers to perform tasks that can be done remotely, such as computer programming, graphic design, data entry, research and writing. 
Markets differ in their scope and focus, but common services provided by the platforms include maintaining job listings, hosting user profile pages, arbitrating disputes, certifying worker skills and maintaining feedback systems.

Although the context is online, many of the same market features found in conventional labor markets also exist in this market.
Employers post job descriptions which workers can search for and apply to.
Employers can assess candidates through interviews and negotiate over wages.
In online labor markets more generally, there are still substantial search frictions \citep{horton2019buyer, horton2017effects}, barriers to entry \citep{stanton2016landing,pallais2010inefficient}, information asymmetries \citep{benson2019can}, and even compelling evidence of employer monopsony power \citep{dube2020monopsony}.

Several marketplace features are worth noting. 
The platform can observe nearly all job search behavior:
it observes which job openings job-seekers apply to, and at what terms; it also observes whether the employer interviewed the applicant and ultimately hired them, forming a contract.\footnote{
  Observing job search effort and intensity is fairly rare, though there are exceptions.
  \cite{mukoyama2018job} examine job search intensity over the business cycle using the ATUS.
  Typically, job search intensity is only seen indirectly, such as through reduced unemployment duration, as in \cite{woodbury1987bonuses} in response to a time-limited bonus for finding a new job.
  \cite{adams2018minimum} does find some increases in search intensity among those already looking for work in response to minimum wage increases. 
}
If a contract is formed, the platform observes how many hours are worked, and at what wage.
Although some workers and firms do haggle, most employers seem to take worker wage bids as take-it-or-leave-it offers \citep{barach2020}.
There is no collective bargaining or even channels for workers to confer with each other---factors that might otherwise lead to sticky wages, as in \cite{saez2019payroll}. 

To work on hourly contracts, workers use a kind of digital punch clock, and so hours are measured essentially without error.
Hourly contracts are the focus on this market, though fixed price contracts are allowed.
For this paper, I exclusively focus on hourly contracts.
For hourly contracts, the hired worker is free to bill hours as they see fit, though employers can cap hours-per-week \emph{ex ante} and can file a dispute if they object to the quality of the work.
Disputes are fairly rare, as employers can observe worker effort in real-time through platform-mandated screen capture software and offer guidance or criticism. 

In addition to market wages and wage bids, the platform can also observe a worker's ``profile wage'' or the wage they list on their worker profiles, which is a kind of online resume.
Employers view these profiles when deciding whom to recruit, and so it tends to be close to what a worker is actually willing to accept.

In terms of market composition, most employers are from the US and the remainder of the demand side are typically from other high-income countries.
The supply side is far more diverse.
The platform knows which country each worker is from, and all market participants can observe the country of every other market participant (who also know this). 
Regardless of which countries the employer and worker are from, all contracts are denominated in US dollars.

Earnings from contracts are kept in account for the worker.
Workers are free to keep their earnings with the platform, but the account is not interest bearing.
This gives workers an incentive to frequently transfer from the platform to their home country bank---at least for workers from countries where transfer fees are \emph{ad valorem} and with a minimum fixed component.
The platform transfers earned money to the worker's bank when the worker requests it.
The worker's bank then does the currency conversion. 
My understanding is that at the time of the ruble collapse, Russian workers faced small transactions costs in transferring their earnings from the platform to Russian banks. 

Prior to the Russian financial crisis, the exchange rate between the ruble and the dollar had been fairly stable since the great recession. 
The ruble began to fall in earnest in the middle of 2014.
Although I will plot time price over time, to give a sense of the fall, in the \nth{2} week of 2014, one US dollar cost about 33 rubles; by the \nth{2} week of 2015, one US dollar cost about 63 rubles at official Bank of Russia exchange rates.

Although much of the platform literature has focused on platform competition, the Russian outside option was not necessarily of being on some other two-sided platform, but simply working in the larger labor market or enjoying leisure.
This is likely commonplace in many online marketplaces where there is still a substantial offline counterpoint;
people still find spouses, drive taxis, walk dogs, assemble furniture, and so on outside of a computer-mediated platform context.

\subsection{Conceptual issues} \label{sec:cf} 
Before getting into the details of how Russians on the platform behaved, it is useful to consider what can be learned from this context. 
The Russian financial crisis changed many things that could potentially have had an independent effect on Russians.
These other factors could cause an under- or over-statement of the effects of changes in the ruble price.
Furthermore, we need a framework for interpreting what we observe in terms of economic theory. 

The claim that the ruble decline increased the real returns to on-platform work deserves greater scrutiny. 
For earning dollars to become more attractive to Russians, we need to assume that Russians were mostly consuming in rubles.
This seems likely.
But we also need to know about the price level in Russia, both for consumption and in the labor market. 
There was substantial inflation in goods prices in Russia \citep{liefert2019}.
I have no good evidence on Russian home country wages for my population---which is mainly software developers---but overall, in 2014 and 2015, inflation was overcoming any nominal wage growth and Russian real wages were likely declining, at least on average.\footnote{
  ``How falling wages are squeezing Russian households,'' World Economic Forum, \href{https://www.weforum.org/agenda/2015/08/how-falling-wages-are-squeezing-russian-households/}{https://www.weforum.org/agenda/2015/08/how-falling-wages-are-squeezing-russian-households/} Accessed Online: 3/29/2020. 
}

It seems likely working online was becoming more desirable for Russians.
How much more desirable is hard to say, but the dramatic increase in Russian participation makes the sign of the effect clear. 
Furthermore, the close tracking of the ruble price and the Russian response (which I will show) suggests it was primarily what Russians were responding to---or they were responding to something that was moving more or less in lockstep with the ruble. 
All these issues aside, my interest is less in precisely estimating various labor supply elasticities but more in seeing what steps Russians took to act on their changed preferences, and how these decisions affected outcomes, without caring too much why those preferences changed, or by how much. 

It is important to note that even if the ruble collapse likely changed the real returns to working, it only had the \emph{potential} to change market wages.
The effects on realized wages depends on how Russian workers changed their wage bidding, if at all, and how employers reacted to those wage bids.
This point is not typically applicable in the labor supply literature, which starts from the assumption that workers are price-takers with respect to the wage and are costlessly adjusting hours-worked in response.\footnote{
  Obviously, many workers on conventional markets face hours restrictions \citep{ham2002testing}, but the competitive perspective has workers choosing employers offering different hours-worked bundles or even flexibility in light of their preferences, even if it takes some time to act on those preferences \citep{kuhn2004monopsony}.
  There is substantial evidence that at least some workers value flexibility in hours-worked \citep{chen2019value}, though how widespread and strong these preferences is less clear \citep{mas2017valuing}.
}

Assuming we can treat the ruble collapse as a \emph{de facto} wage shock, there are obstacles if the goal were to estimate a life-cycle model of labor supply. 
We do not know what Russians believed about the time courses of the ruble, nor do we know their wealth, consumption, non-labor sources of income, or what borrowing constraints they faced. 
That being said, given that the proximate cause was geopolitical events that could in principle be resolved quickly, it seems likely that Russians viewed the higher real wage opportunity as unexpected and temporary.
Furthermore, if Russians believed in long-run purchase price parity, they would expect this change to be temporary, though given how persistent gaps exist between nominal and real exchange rate fluctuations, how ``temporary'' is not clear. 
Even if the course of the ruble was unclear, Russians surely know about the changes, unlike, for example, workers needing to learn about UI benefit generosity before being affected by them \citep{lemieux2000supply}. 
As a transient change observed more or less immediately by Russians, the elasticity estimated in a panel is arguably ``Frisch-like'' even though we lack some of the right measurements for this claim \citep{macurdy1981empirical}.

\section{The ``macro'' view of the ruble collapse} \label{sec:country_results} 
To begin, I compare country-level outcomes over time in a panel: numbers of active workers, the number of applications sent, the number of hires made, the average wage bid, and so on.
I compare Russia to non-Russian countries that experienced relatively small currency fluctuations during the period covered by the data.

\subsection{Panel definition}
I use \numCountries{} different countries as the comparison units, with Russia included.
These are the largest participating countries by worker count, excluding Ukraine which was removed because it was also affected directly by the Russian financial crisis.\footnote{
  Some other countries were removed because of data quality issues. 
} 
I construct a balanced weekly panel, with \numWeeks{} weeks per country, for a total of \numObs{} observations.
The panel starts on \firstWeek{} and ends on \lastWeek{}, which is where my data end.
This panel is constructed from all applications sent during the period covered by the panel.

\begin{table}
  \caption{Country-level measures in the \nth{2} week of 2014 and 2015, during which ruble lost approximately half its value against the US dollar} \label{tab:summary_stats_panel}
  \begin{small}
    \captionsetup[table]{labelformat=empty,skip=1pt}
\begin{longtable}{rrrrrrr}
\toprule
& & & \multicolumn{2}{c}{Means} & & \\ 
 \cmidrule(lr){4-5}
Year & Active & Apps/Active & Wage (log) & Win Rate & Hires & Rus-Index \\ 

\multicolumn{1}{l}{Philippines} \\ 

2014 & $10,204$ & $5.77$ & $1.49$ & $0.029$ & $1,703$ & $0.007$ \\ 
2015 & $9,615$ & $5.71$ & $1.67$ & $0.026$ & $1,413$ & - \\ 

\multicolumn{1}{l}{Kenya} \\ 

2014 & $693$ & $5.34$ & $1.82$ & $0.033$ & $121$ & $0.008$ \\ 
2015 & $776$ & $5.10$ & $2.01$ & $0.027$ & $106$ & - \\ 

\multicolumn{1}{l}{Bangladesh} \\ 

2014 & $9,748$ & $10.43$ & $1.17$ & $0.013$ & $1,354$ & $0.008$ \\ 
2015 & $9,392$ & $9.45$ & $1.54$ & $0.013$ & $1,180$ & - \\ 

\multicolumn{1}{l}{United States} \\ 

2014 & $5,241$ & $4.36$ & $2.75$ & $0.040$ & $919$ & $0.014$ \\ 
2015 & $5,149$ & $4.45$ & $2.82$ & $0.039$ & $886$ & - \\ 

\multicolumn{1}{l}{United Kingdom} \\ 

2014 & $877$ & $4.69$ & $2.64$ & $0.030$ & $124$ & $0.016$ \\ 
2015 & $981$ & $4.80$ & $2.79$ & $0.041$ & $195$ & - \\ 

\multicolumn{1}{l}{Pakistan} \\ 

2014 & $5,196$ & $7.56$ & $1.78$ & $0.020$ & $795$ & $0.018$ \\ 
2015 & $5,117$ & $7.69$ & $1.98$ & $0.018$ & $706$ & - \\ 

\multicolumn{1}{l}{Canada} \\ 

2014 & $709$ & $5.20$ & $2.65$ & $0.039$ & $144$ & $0.020$ \\ 
2015 & $755$ & $5.51$ & $2.84$ & $0.038$ & $160$ & - \\ 

\multicolumn{1}{l}{India} \\ 

2014 & $19,078$ & $8.54$ & $2.09$ & $0.011$ & $1,873$ & $0.022$ \\ 
2015 & $21,454$ & $9.15$ & $2.17$ & $0.009$ & $1,771$ & - \\ 

\multicolumn{1}{l}{Russia} \\ 

2014 & $623$ & $5.51$ & $3.00$ & $0.027$ & $92$ & - \\ 
2015 & $1,480$ & $6.01$ & $2.96$ & $0.026$ & $234$ & - \\ 
\bottomrule
\end{longtable}

    \emph{Notes:} Country-level comparisons of activity in an online labor market across the second week of January, in 2014 and 2015.
    The column labeled ``Rus-Index'' is measure of how closely workers from a country compete with Russians, on average prior to the ruble collapse. 
    \end{small}
\end{table}

Table~\ref{tab:summary_stats_panel} presents data from the panel from two points in time: the \nth{2} week of 2014 and the \nth{2} week of 2015.
I use the \nth{2} week to avoid the New Year's holiday week.
Comparing the Russia rows, we can see a more than 3x increase in the number of active Russians, where ``active'' is defined as sending at least one job application.
Most other countries show no large changes.
There is also an increase in applications per active Russian worker, though other countries also show changes of similar or greater magnitude.
The Russian win rate---the probability they are hired for a given application---does not change.

The Russian average wage bid is slightly lower at the end of the period covered by the panel.
Some countries experience large wage increases over the time period, particularly countries that started with relatively low wages at the start of the panel
This was due to a minimum wage change on the platform, which I will discuss at length.

The number of Russians hired per week increased substantially, going from \NumRusHiredWeekFirst{} to \NumRusHiredWeekLast{}, while other countries show no changes nearly as large.
Some of the lower wage countries show fairly large declines; some of the higher wage countries show increases.
This is also likely the result of the minimum wage, as it caused a substantial degree of labor-labor substitution towards relatively high-wage workers. 

In Table~\ref{tab:summary_stats_panel}, countries are ordered from least to most similar to Russian in terms of application overlap i.e., workers from the Philippines tend to work in very different categories from Russians, while workers from India compete more directly, on average.
However, a comparison of summary statistics for India and Russia show that they are not that similar on many dimensions---Russians have far higher wages, send fewer applications and have nearly double the per-application win rate.
In terms of application behavior and outcomes, Russians are more similar to workers from countries like Canada and the United States.

\subsection{Market quantities over time}
\begin{figure}
\caption{The ruble collapse and weekly country-level market outcomes, de-meaned} 
\label{fig:country_level_outcomes}
\centering
\begin{minipage}{1.0 \linewidth}
  \includegraphics[width = \linewidth]{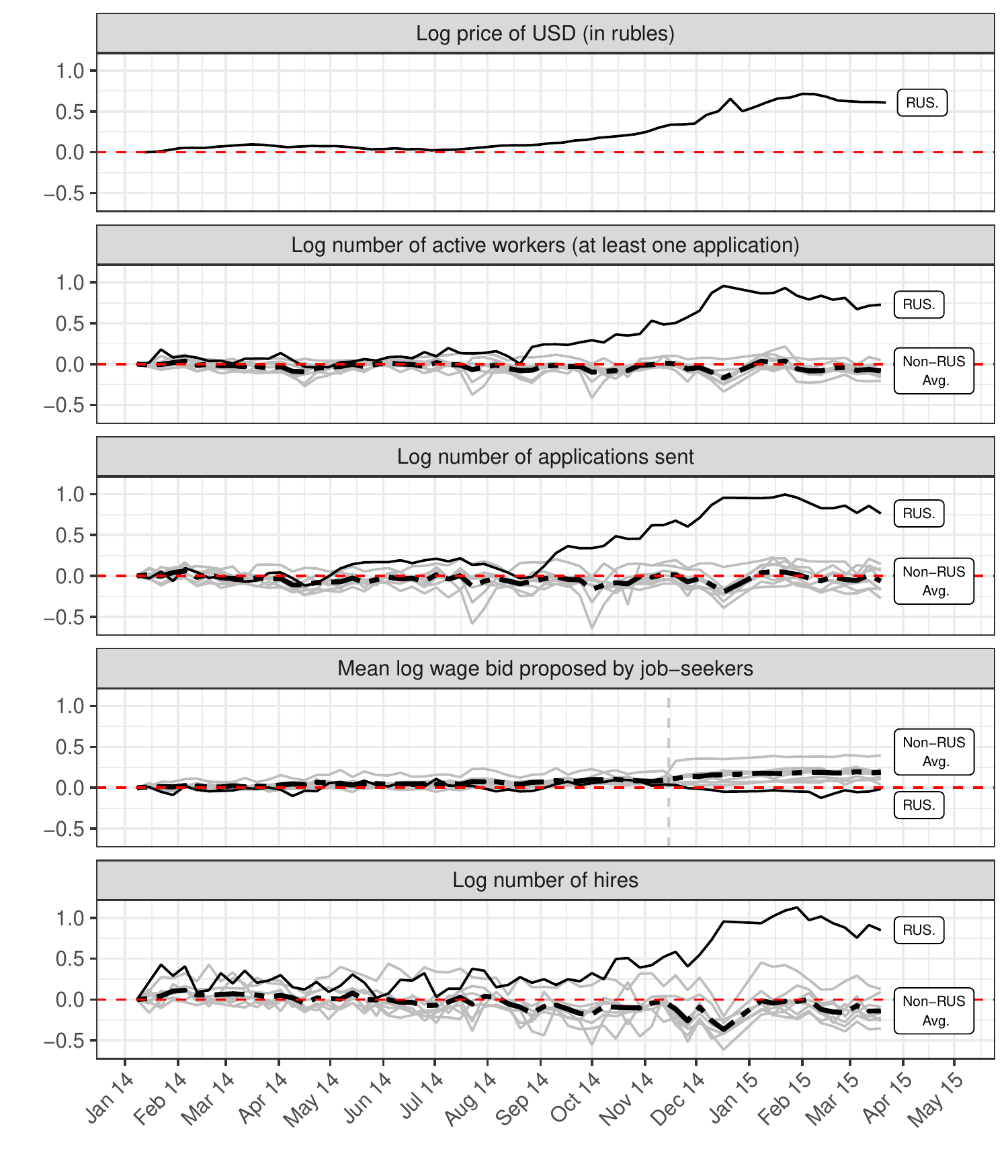} \\
  \begin{footnotesize}
    \begin{singlespace}
      \emph{Notes:} This figure shows the time series of a collection of country-level outcomes.
      All series are de-meaned to the first period.
      The non-Russian countries are the same as those listed in Table~\ref{tab:summary_stats_panel}. 
    \end{singlespace}
    \end{footnotesize}
\end{minipage}
\end{figure}

The by-week log price of one US dollar in rubles, is plotted in the top facet of Figure~\ref{fig:country_level_outcomes}.
Each facet below the ruble facet shows the time series for country-level outcomes.
All time series are de-meaned to 0 at the start of the panel. 
Russia is a heavy black line in shape, while all other countries are light gray lines.
The average for all non-Russian countries is a heavy dashed line. 

In the second facet from the top, labeled ``Log number of active workers (at least one application),'' we can see the dramatic increase in Russians engaged in job search on the platform.
For Russia, the log number active line tracks the log US dollar in ruble line, but with a larger magnitude.
The number active line tops out at nearly 1, despite the US dollar-to-ruble line topping out at about 0.75, giving an eyeballed elasticity greater than 1. 

The third facet from the top, labeled ``Log number of applications sent'' shows a large increase in applications from Russians.
The curve closely tracks the number of active workers, not just in shape but also in magnitude.
An implication of the ``active'' and ``applications'' curves having the same magnitude is that there is no evidence of an intensive margin effect (i.e., number of applications send conditional upon sending any).
However, it is important to remember that given the large extensive margin effects, an aggregate intensive margin calculation is likely masking important composition changes.
The kinds of Russians joining the platform in response to the ruble collapse might have relatively little capacity to take on more work, and thus send relatively few applications compared to ``incumbent'' Russians already active.
To explore these issues, I will turn to an individual worker panel analysis in Section~\ref{sec:individual_results}.

The facet labeled ``Mean log wage bid proposed by job-seekers'' we can see that the Russian line eventually dips below 0, but eventually returns to 0 at the end of the panel.
However, other countries go up: in the light gray lines, we can see that he mean wage for some other countries shows a substantial and sudden increase indicated at the vertical dashed line.
As mentioned earlier, the cause is the imposition of a platform-wide minimum wage of \$3/hour in mid-November 2014, which strongly increased average wages in several low income countries \citep{horton2017price}. 
This platform wide minimum wage will require us to take care when doing a difference-in-differences analysis.

Prior to the minimum wage imposition, note that the Russian average is right with the other countries, despite there already being enormous Russian increases in market participation by then.
This is suggestive evidence of little pass-through of the ruble decline into wage bids.
A gap emerges later between Russia and other countries, but to the extent there is a real relative Russian decline, it could be a composition effect---new entrants could be less productive---or a market effect, with increased competition in Russian-heavy categories.

The outcome in the bottom facet of Figure~\ref{fig:country_level_outcomes} is the ``Log number of hires.'' 
For Russians, we can see hires curve closely tracks the applications curve and has a similar magnitude. 
The similarity in magnitude implies no change in per application win rates---the added Russian applications seemingly convert into hires at the same rate as before. 

The count of hires is not the same as the count of hours-worked.
However, the size of contracts won (in terms of hours-worked) did not change on average, and so this hires elasticity is likely equivalent to the hours-worked elasticity, though there are some conceptual issues with this claim, and labor supply is better investigated at the micro level.\footnote{
  I could plot hours-worked here, but there is an issue of whether to allocate those hours to the date the contract was formed, or the date the hour was actually worked.
  As much contracts are short in duration, this not likely to create much of a difference empirically. 
}

\subsection{Russian work specialization} \label{sec:russian_focus}  
\begin{figure}
\caption{Changes in the fraction of applicants from Russia} \label{fig:composition_change}
\centering
\begin{minipage}{0.95 \linewidth} 
\includegraphics[width = \linewidth]{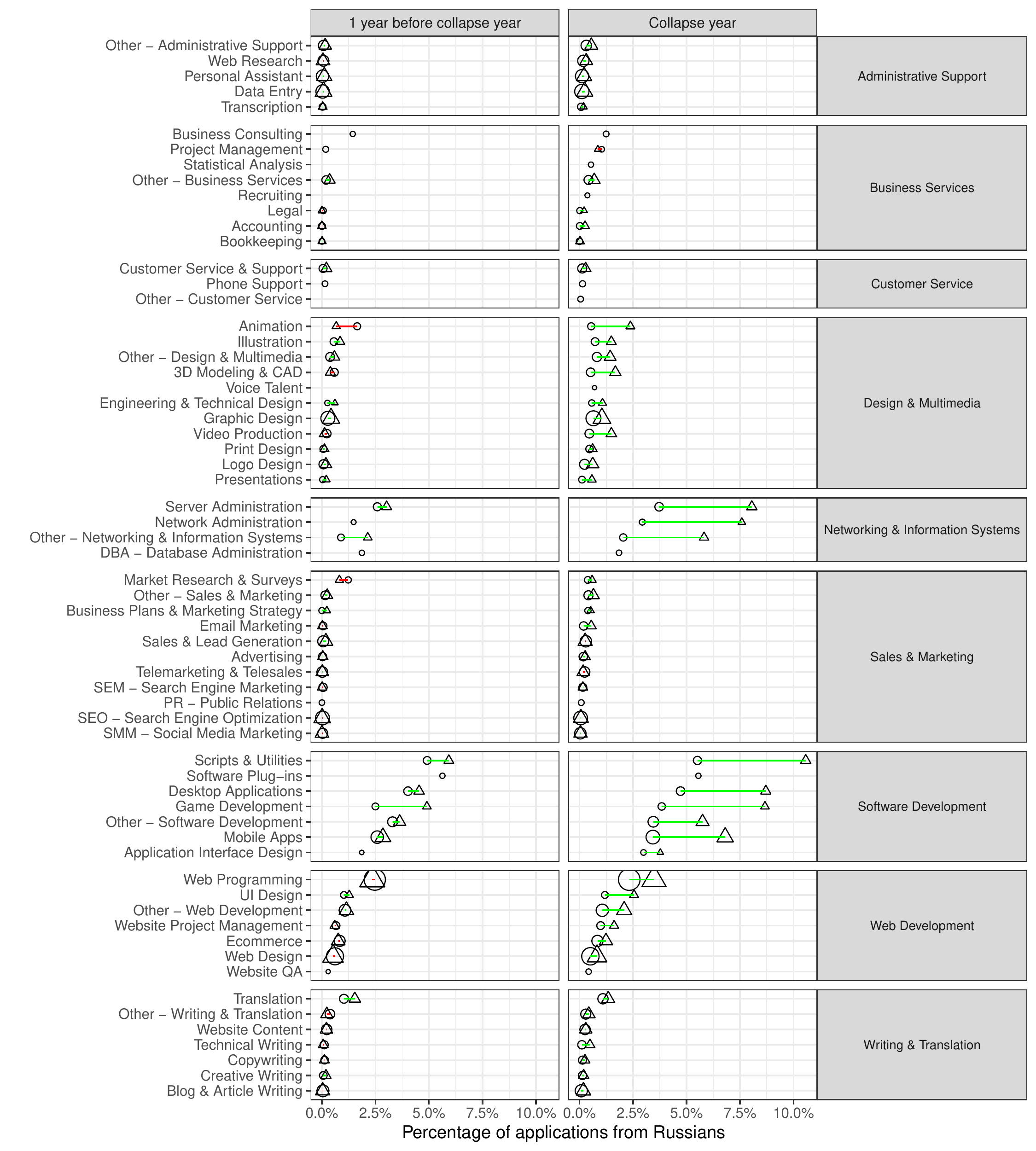} \\
\emph{Notes:}
     {\footnotesize This figure shows fraction of applicants from Russia, per the platform sub-category. 
      The right panel, labeled ``collapse year'', shows the fraction of applications during two periods:  
The pre-period is April 1, 2014 to July 1st, 2014, which is indicated by a circle, and a post period from 
January 1st, 2015 to March 1st, 2015, which is indicated by triangle.
The left panel, labeled ``1 year before collapse year'' is the same as the right panel, but with the observations shifted back one year. 
Point sizes are scaled by the number of total job openings in that sub-category.
The sample is restricted to categories with at least 300 total openings in both pre- and post periods for the collapse year and the year before the collapse.}     
\end{minipage}
\end{figure}

The increase in the number of applications from Russians was not evenly spread across all kinds of work.
Figure~\ref{fig:composition_change}, shows how the fraction of applicants from Russia changed over time.  
The right panel, labeled ``collapse year'', shows the fraction of applications during two periods:  
The pre-period is April 1, 2014 to July 1st, 2014, which is indicated by a circle, and a post period from 
January 1st, 2015 to March 1st, 2015, which is indicated by triangle.
The left panel, labeled ``1 year before collapse year'' is the same as the right panel, but with the observations shifted back one year. 
In collapse year, in both the pre- and post-periods (in the right panel) and the same comparisons pushed one calendar year in the past (in the left panel). 
A circle corresponds to the ``pre'' period and the triangle to the ``post'' period.

We can see that in certain categories and sub-categories, the collapse of the ruble lead to large increases in the fraction of applications coming from Russians, whereas in other areas, there was minimal impact.
The plot makes it clear that the influx of Russians was concentrated in categories of work that Russians already focused on before the ruble collapse. 

\subsection{Accounting for competition among countries and the minimum wage issue}
One limitation of the graphical approach of Figure~\ref{fig:country_level_outcomes} is that it does not account for the fact that some countries are closer competitors with Russians than others.\footnote{
  Coincidentally, this ``ethnic specialization'' design mirrors the design of \cite{borjas2012collapse}, which relied on a previous Russian economic crisis to estimate the effects on American mathematicians that were in Russian-heavy sub-fields of mathematics. 
}
A sudden influx of Russians affected certain kinds of work---namely work that Russians tend to specialize in---which in turn could affected certain workers---namely non-Russians that tend to focus on the same kinds of work as Russians.
This could make some countries bad controls.

To create a measure of Russian competition exposure, I start with each application sent by any worker from any country and compute the fraction of their fellow applicants to that job opening that were Russian.
I then average this measure over all workers in a particular country.
This is done with applications prior to the start of the panel.
In a regression, I can then interact this measure with the ruble price, allowing close competitors to Russians to be more affected.

I can do something similar for the minimum wage, by computing the lowest pre-minimum wage average wage per country and interacting this with a post minimum wage imposition indicator.
This allows countries that were affected by the minimum wage to have a different response after the minimum wage imposition. 
As it turns out, the competition/overlap issue seems to be more of a problem in theory than in practice;
the minimum wage issue matters more. 

\subsection{Country-level regression evidence} 
To address the competition/overlap and the platform minimum wage issues, I switch to a regression framework.
My preferred specification is: 
\begin{align} \label{eq:country_panel}
  y_{ct} &= \beta_1 \left( \log p_t \times \textsc{Russian}_c \right) + \\
  &\beta_2 \left( \log p_t \times RI_c \right) + \beta_3 \left(\textsc{Post}_t \times \underline{w}_c\right) + \delta_t + \gamma_c + \epsilon_c \nonumber,  
\end{align} 
where $y_{ct}$ is some outcome of interest, such as log applications sent per week by workers from country $c$ during week $t$, $p_t$ is the price of one US dollar in rubles at the start of week $t$, $\textsc{Russian}_c$ is an indicator for whether the observed country $c$ is Russia, $RI_c$ is the Russian competition index (set to 0 for Russia), $\underline{w}_c$ is the lowest observed log wage for a country $c$, $\textsc{Post}_t$ is an indicator for periods after the imposition of the minimum wage, and finally, $\delta_t$ and $\gamma_c$ are week- and country-specific fixed effects.
I report four different specifications:
\begin{itemize}
\item Russian Index Interaction and Min Wage Interaction (full Equation~\ref{eq:country_panel})
\item No Interaction ($\beta_2 := 0$ and $\beta_3 := 0$). 
\item Min Wage Interaction only ($\beta_2 := 0$).
\item Russian Index Interaction only ($\beta_3 := 0$).
\end{itemize}

\begin{figure}
\caption{Effect of ruble price changes using a country-week panel} \label{fig:country_panel_results}
\centering
\begin{minipage}{1.0 \linewidth}
  \includegraphics[width = \linewidth]{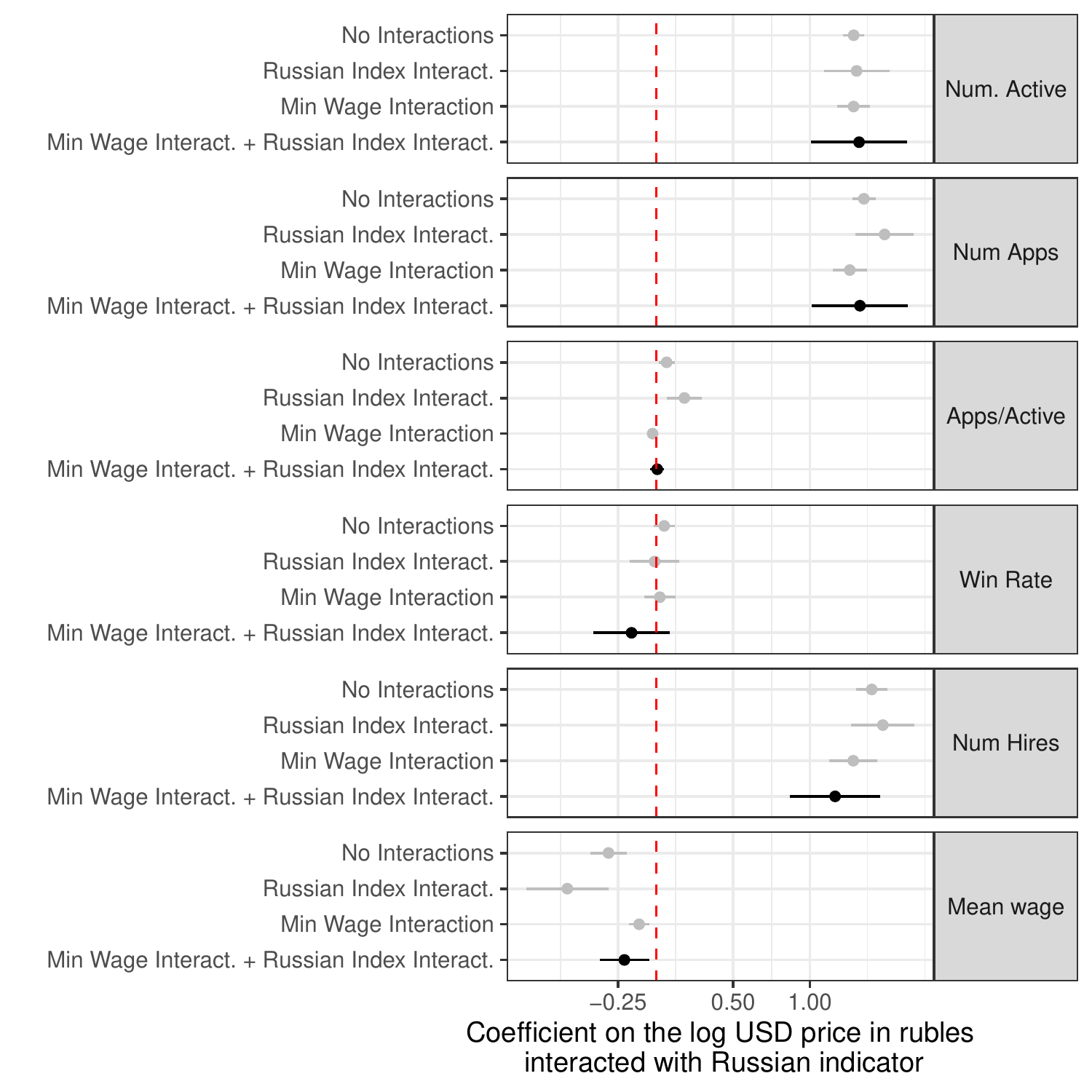} \\
  \begin{footnotesize}
    \begin{singlespace}
      \emph{Notes:} This panel reports estimates of Equation~\ref{eq:country_panel}.
      These results in table form are in Appendix~\ref{sec:table_versions}. 
    \end{singlespace}
    \end{footnotesize}
\end{minipage}
\end{figure}

Figure~\ref{fig:country_panel_results} reports $\beta_1$ from Equation~\ref{eq:country_panel}, with each facet reporting the effect on a different outcome. 
All of the specifications are shown; my preferred specification is in black, while the others are in gray.\footnote{
  The full regressions in table form are in Appendix~\ref{sec:table_versions}.
}
Standard errors are clustered at the level of the country.\footnote{
  To check for \cite{bertrand2004much} problems, I also did block bootstrap on the country (leaving Russia in every sample).
  The bootstrap standard errors were nearly identical, and so I report the conventional standard errors with the appropriate clustering. 
}
In the top facet, the outcome is the log number of active workers.
The point estimates are similar regardless of the specification: 
the number of active Russians increased dramatically, with a 10\% associated with about a \AppActiveElasticityTenPct{}\% increase in active Russians.

In the next facet, the outcome is the log number of applications.
Again, the precise specification does not seem to matter much, in that the effects are similar to what we saw with the number of active Russians:
10\% increase in the value of a US dollar in rubles is associated with \AppElasticityTenPct{}\% more applications from Russians.

Given that the effects of the ruble decline has similar percentage effects on the number of active workers and the total number of applications, we expect little effect on the intensive margin with respect to applications.
In the third facet from the top, the outcome is the log of the number of applications sent divided by the number of workers active.
Given the large extensive margin effect relative to the overall increase, there is not much room for an intensive margin effect among Russians.
Indeed, the point estimates are all close to zero. 
As mentioned earlier, this country-level regression is not capturing an intensive margin effect---I will turn to this issue in the micro analysis.

The third facet from the bottom shows the effect on win rates. 
All the specifications are close to zero, implying no change.
As such, we should expect the total number of hires to be proportional to the number of applications.
In the second panel from the bottom, we see that prediction is borne out:
the number of Russian hires increases dramatically, with elasticity point estimates similar to what we observed for applications and active workers. 

In the bottom facet, the outcome is the mean log wage bid. 
For this outcome, the specification matters. 
Although all point estimates are negative, the two specifications that have the minimum wage interaction term have estimates closer to zero. 
As the minimum wage imposition occurred when the ruble was near its minimum, the inclusion of minimum-wage affected countries (without $\beta_3 := 0$) makes Russian wages look worse when this effect is not captured with the appropriate specification. 

\section{The ``micro'' view of the ruble collapse} \label{sec:individual_results} 

I now take an individual worker perspective, using only workers that were active pre-collapse, thus fixing the composition of the panel. 
To avoid the complexities associated with the minimum wage policy change (and the unnatural ``exit'' this created for at least some low-wage workers), I restrict the countries to: \CountryList{}.
These are higher wage countries that are also in somewhat less competition with Russian workers.

To define the sample, I take all workers that were active---defined as sending at least one job application---from \SampleStart{} to \SampleEnd{}. 
There are a total of \SampleSize{} unique workers, of whom \SampleSizeRussians{} are Russian.
Mirroring the country-level panel, the individual level panel covers \PanelStartIndv{} to \PanelEndIndv{}.
There are \numPeriods{} one week periods, giving a panel of \numObsInd{} worker-week observations.\footnote{
 It is slightly larger that the country panel because of slight differences in panel construction.
}

Given the lack of importance of the ``overlap'' measures and the exclusion of minimum-wage affected countries, I use a simpler specification: 
\begin{align} \label{eq:individual}
   y_{it} = \beta_1 \left( \log p_t \times \textsc{Russian}_i \right) + \delta_t + \gamma_i + \epsilon_i, 
\end{align} 
where $y_{it}$ is some weekly outcome for worker $i$ in week $t$, $\textsc{Russian}_{i}$ is an indicator for the worker being from Russia, $p_t$ is the average price of one US dollar in rubles during the week $t$, $\delta_t$ is a week-specific fixed effect and $\gamma_i$ is a worker-specific fixed effect.
Standard errors are clustered at the level of the individual worker.

The first outcome of interest is the extensive margin, or whether the worker was active on the platform in the sense of earning some amount of money.
Unlike the other individual outcomes, this cannot be estimated using a log specification, and the change has to be compared relative to some baseline to construct an elasticity. 
Using the Russian baseline probability of working some number of hours in the first 4 weeks of the start of the panel, the implied extensive margin elasticity with respect to the log price of the US dollar in rubles is \EffectActiveExtensive{}. 

\begin{figure}
\caption{Individual-level estimates of effects of ruble prices changes on individual outcomes} \label{fig:individual_panel_results}
\centering
\begin{minipage}{0.95 \linewidth} 
\includegraphics[width = \linewidth]{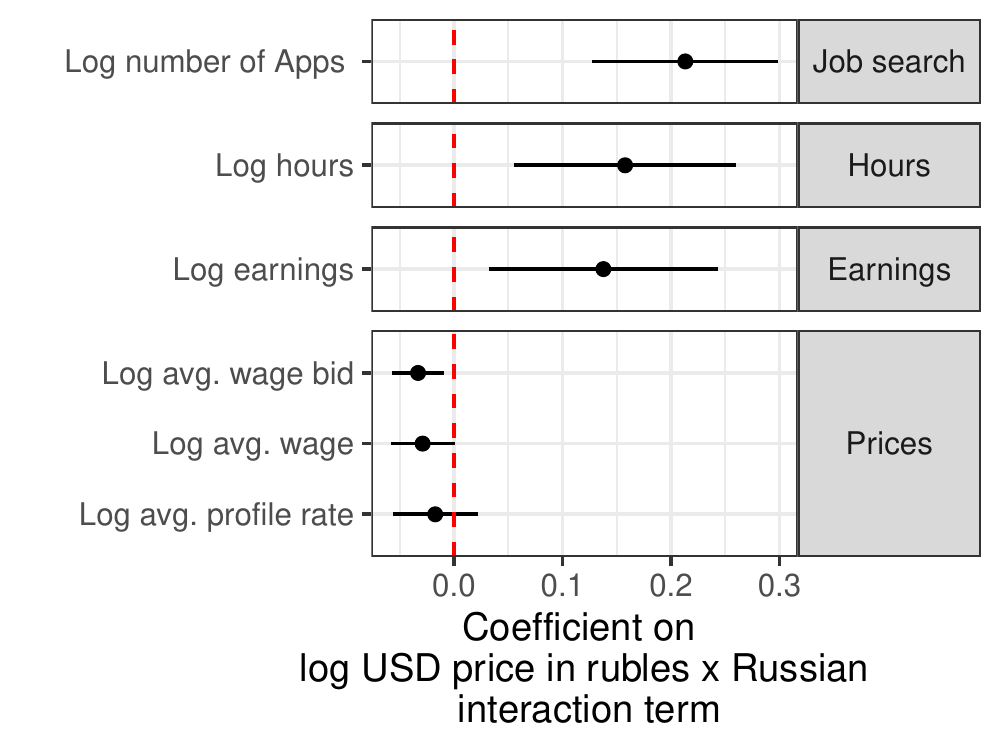} \\
\emph{Notes:} This figure reports estimates of $\hat{\beta}_1$ from Equation~\ref{eq:individual}.
      These results in table form are in Appendix~\ref{sec:table_versions}. 
     {\footnotesize }
\end{minipage} 
\end{figure}

For the rest of the outcomes, the sample is conditioned on the worker being either working some number of hours and/or sending an application. 
These estimates of $\beta_1$ from Equation~\ref{eq:individual} are presented in Figure~\ref{fig:individual_panel_results}.

In the top facet of Figure~\ref{fig:individual_panel_results}, the outcome is the log number of applications sent by a worker per week, conditional upon sending any.
The outcome, in levels, has summary statistics \AppsSScop{}. 
The implied elasticity with respect to the US dollar price in rubles is \EffectAppElasticity{}. 
This is considerably smaller than the application macro estimate---which included extensive and intensive margin effects---but it is also non-zero, whereas in the macro panel, there was no evidence of an intensive margin increase in applications.

In the second facet from the top, the outcome is the log number of hours-worked by a worker per week, conditional upon any.\footnote{
  Manual hours are possible, giving implausibly high values. I cap hours-per-week at \MaxHours{}, replacing those entries with \MaxHours{} exactly.
  This affects less than 1\% of observations in the panel.
}
This outcome, in levels, has summary statistics \HoursSScop{}.
The hours-worked elasticity is \EffectHoursElasticity{}, which is similar to the application intensity elasticity.

The outcome in the third facet from the top is the log earnings by a worker per week, conditional upon any.
The estimated elasticity is also similar in magnitude to the hours-worked and application-intensity elasticities.
That the effects on hours and earnings are similar foreshadows the effect on wages---namely that there is little effect. 

For wages, I report the effect on three outcomes: the log average wage bid, the log average wage worked at, and the log average profile rate.
Summary statistics for the actually worked-at wages are, in levels, are \WageSScop{}. 
There is some slight evidence of decline in the average wage and a decline in the wage bid; the profile wage is unchanged.
The effect on the individual wage estimate is negative and statistically significant, but fairly close to zero: the implied elasticity is just \EffectWageElasticity{}.
Note that this estimate is about 1/10th the size of the macro panel estimate, with the difference likely explained by composition effects.

There is no evidence that workers lowered their wage bids by an economically significant amount to obtain more hours of work.
If we take the bargained wage view, this results implies workers have nearly all the bargaining power.
However, a more sensible interpretation is that job-seekers and employers act as price-takers and a bargaining lens is the wrong lens.

\subsection{Intensive margin labor supply elasticity by current hours-per-week}
Given the large extensive margin effects revealed in the macro panel, it is useful to consider whether Russians lightly attached to the market pre-collapse were particularly elastic. 
A stylized fact from the labor supply literature is that workers who are less ``attached'' to the market tend to be more elastic \citep{blundell1999labor}.
Adapting this to our setting, it seems likely that incumbent Russian workers would respond differently depending on how many hours they were already working pre-collapse.

To assess whether intensive margin elasticities vary by pre-collapse hours-worked, I compute the fraction of all weeks that workers were working at least some number of hours in the 6 months prior to the start of the panel.
Note that this sample is defined by actually working, not just sending an application. 
For weeks where they were active, I compute the average hours per week, and I compute the fraction of weeks active.
I then divide workers into quartiles based on the their average hours per week, conditional upon working. 

\begin{figure}
  \caption{Hours-worked and estimated intensive margin elasticities for workers active before the ruble collapse} \label{fig:hours_by_band}
\centering
\begin{minipage}{0.8 \linewidth}
  \includegraphics[width = \linewidth]{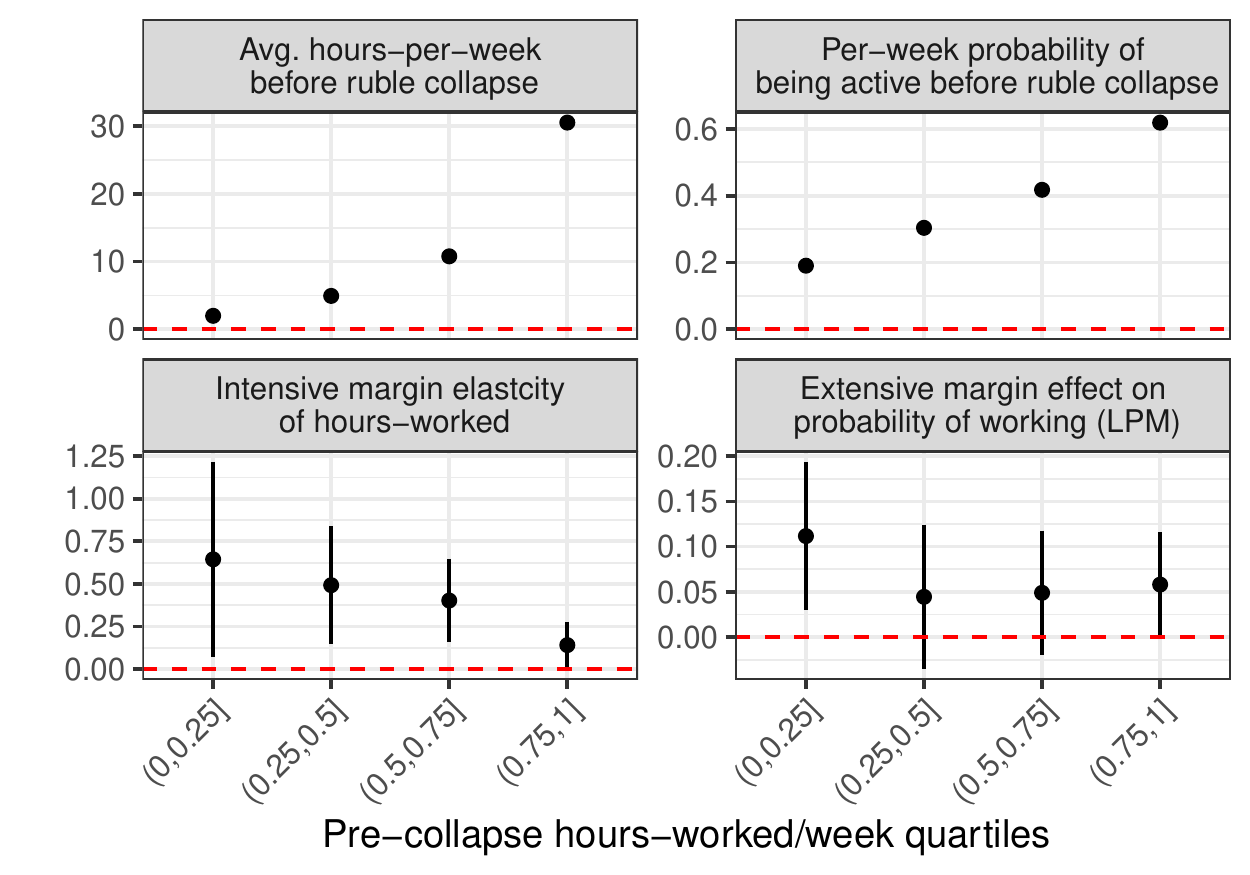} \\
  \begin{footnotesize}
    \begin{singlespace}
      \emph{Notes:} This figure reports characteristics of Russian workers by decline in terms of pre-collapse activity.
      The bottom panels report estimates of $\beta_k$ from Equation~\ref{eq:individual_hours}. 
    \end{singlespace}
    \end{footnotesize}
\end{minipage}
\end{figure}

Figure~\ref{fig:hours_by_band}, in the upper left facet, shows the average hours-worked by quartile pre-collapse.
The highest quartile is working close to full time, on average, while the bottom quartile is only working a few hours per week.
In the top right facet, the per-week probability of being active pre-collapse is reported.
Even in the top quartile it is only 60\%, but recall that in the pre-period, workers are ``joining'' the platform and so some of these non-working weeks were before they had joined.

To estimate effect of the ruble collapse, I use the specification 
\begin{align} \label{eq:individual_hours}
  y_{it} = \sum_g \beta_g \left( \log p_t \times \textsc{Russian}_i \times H_g \right) + \sum_g \delta_{gt} + \gamma_i + \epsilon_i, 
\end{align}
which has the Russian indicator and log US dollar price in rubles interacted with the quartile indicators.
There is now a separate time effect, $\delta_{gt}$, for each quartile, which is indexed by $g$, as well as an individual fixed effect, $\gamma_i$. 
Standard errors are clustered at the level of the individual worker. 

For the probability of being active, in the bottom right facet of Figure~\ref{fig:hours_by_band}, we can see there is an increase across quartiles, with effects largest in the lowest quartile.
However, effects are generally imprecise.  

In the bottom left facet of Figure~\ref{fig:hours_by_band}, the outcome is the log hours-worked, conditional upon any. 
For the intensive margin, we see fairly high point estimates for workers that where weakly attached, but the estimates are highly imprecise. 
The elasticity declines with greater pre-collapse usage, in the top quartile, the point estimate is about 0.12. 

Despite the limitations the setting imposes on the interpretability of the elasticities, if we take them at face value, the results are broadly inline with other Frisch labor supply estimates in \cite{blundell1999labor}, though they are considerably smaller than the estimates in settings where the changes are much more obviously temporary \citep{angrist2017uber,fehr2007workers}.
The finding of heterogeneity in elasticity by labor market attachment is also found in conventional settings.

\section{Evidence of a demand response?} \label{sec:outcomes_for_russian_vacancies} 

Employers posting jobs that would appeal to Russians presumably received more applications and---our panel evidence notwithstanding---perhaps lower wage bids.
This could cause them to post more Russian-friendly jobs or be more likely to fill them.
To see how this affected the employers, we need to switch to the job post as our unit of analysis.
A key complication is that we now need some measure of how appealing a job opening would be to Russian applicants. 

To construct a measure of Russian exposure, I first construct a historical dataset of pre-collapse job openings, recording the fraction of those applicants that were Russian.
Then, I use the full document-term-matrix for the skills required for that opening, as well as other characteristics set by the employer as predictors.
I then use gradient boosting (using the \verb|xgboost| R package) to train a linear model \citep{friedman2000additive,chen2015xgboost}. 
I use the fitted model out of sample to make predictions for all job openings posted during our panel year. 
The predictive model can explain about 25\% of the variance in the realized fraction of Russian applicants in the panel sample.

\begin{figure}
\caption{Comparison of the predicted fraction of Russian applicants (based on job opening characteristics) versus the realized fraction}
\label{fig:job_openings_over_time}
\centering
\begin{minipage}{1.0 \linewidth}
  \includegraphics[width = \linewidth]{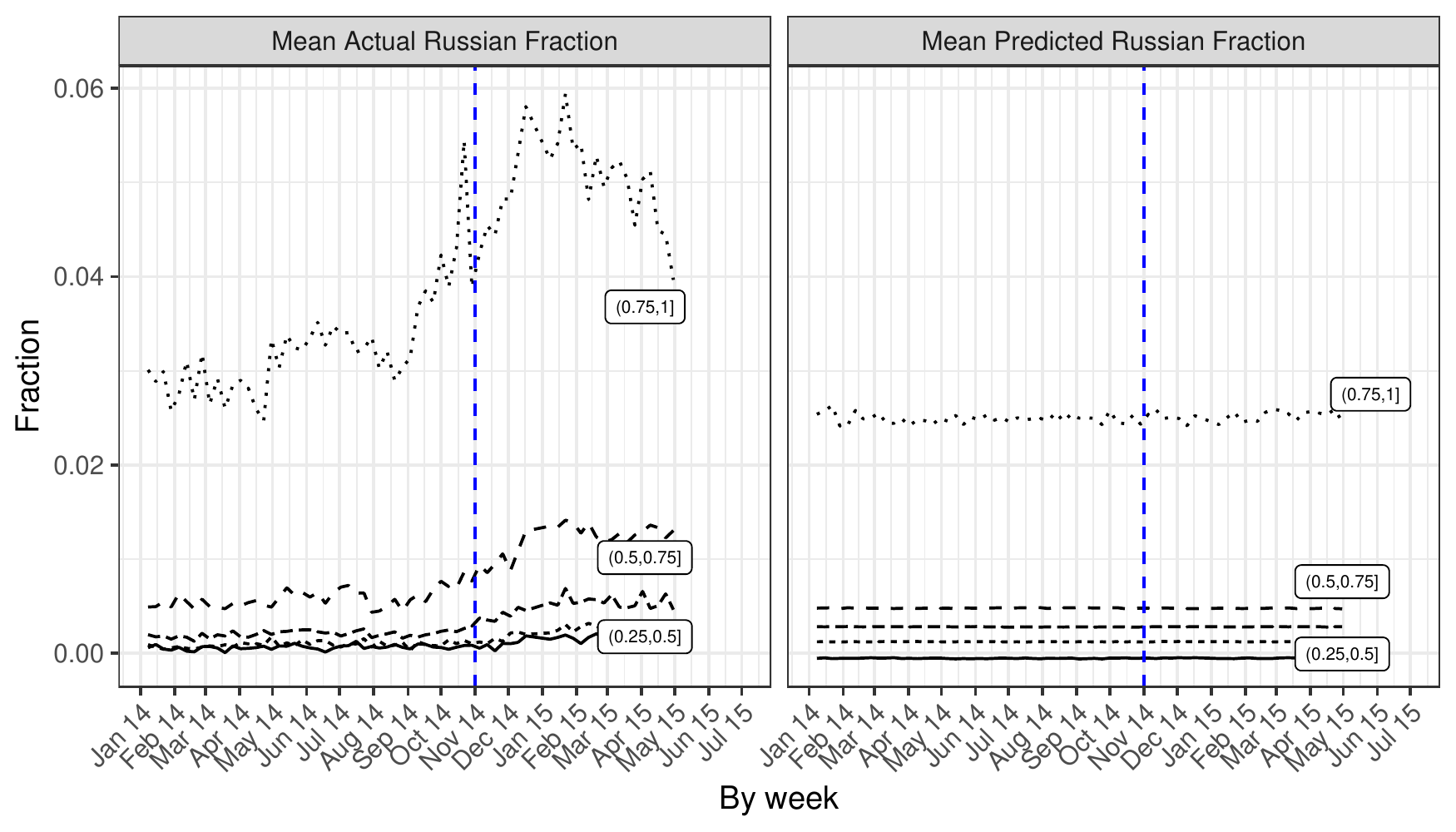} \\
  \begin{footnotesize}
    \begin{singlespace}
      \emph{Notes:}  This figure shows the predicted and actual fraction of Russian applicants for job openings, with  the predictions derived from the text of the job description, skill required and job title.
      The different lines correspond to the different quantiles of the predicted Russian applicant fraction distribution. 
    \end{singlespace}
    \end{footnotesize}
\end{minipage}
\end{figure}

Figure~\ref{fig:job_openings_over_time}, in the right panel, I plot the mean predicted score (the fraction of Russian applicants), by quantile (with the cut points determined by pooling over the entire panel) and by week.
We can see that the predicted score shows no time trend in any level.
This is consistent with there being no demand shift or compositional shift in the kinds of jobs being posted.
If instead more Russian-friendly jobs were posted, these time series would rise as the model would observe more jobs with Russian-friendly characteristics and predict a greater fraction of Russians.
This is not the case. 
In short, there is no evidence employers posted more Russian-compatible job openings. 

In the left panel, I plot the mean actual, realized fraction of Russian applicants, but by predicted quantile.
At first, the predicted and actual match during the period before the ruble began to strongly depreciate.
However, as the ruble depreciates, we can see somewhat of an increase in every quantile---even the lowest---there is an increase in the fraction of actual applicants, though the increase is concentrated among those job openings expected to receive the largest number of Russian applicants.   
Among the top 25\%, the fraction of Russian applicants nearly doubles. 

\subsection{Job opening level outcomes} 
To see how this Russian supply shock affected the openings, I discretize the Russian compatibility score in a collection of $k$ percentile intervals, from lowest score to highest score. 
I then estimate regression of the form 
\begin{align} \label{eq:job_opening_effects}
  y_j & = \sum_k \alpha_k \textsc{RcsInt}_{k(j)} + \sum_k \beta_k \left(\textsc{RcsInt}_{k(j)} \times \log p_{t(j)} \right) + \delta_{t(j)} + \epsilon 
\end{align} 
where $y_j$ is some outcome for job opening $j$, $\textsc{RcsInt}_{k(j)}$ is an indicator for whether the Russian-compatibility score for job opening $j$ is in the $k$th interval, and $p_{t(j)}$ is the price of one US dollar in rubles when opening at time $t$, which is when opening $j$ was posted, and finally, $\delta_t$ is day fixed effect. 

Table~\ref{tab:job_opening_apps} reports the $\hat{\beta}_k$ of the regressions where the outcomes are various job opening level measures.
In Column~(1), the outcome is the log number of Russian applications received, plus 1.
For job openings with relatively lower Russian-compatibility scores, the effect is positive but not large in magnitude.
However, the effect is increasing in the Russian compatibility score.
At the highest band---the most Russian compatible openings---the increase is large and the elasticity is \RussianAppPoolElasticity{}.

\begin{table}[!htbp] \centering 
  \caption{Effects of the ruble collapse on per-vacancy measures of competition, wage bidding, hires and total wage bill} 
  \label{tab:job_opening_apps} 
\footnotesize 
\begin{tabular}{@{\extracolsep{5pt}}lccc} 
\\[-1.8ex]\hline 
\hline \\[-1.8ex] 
 & \multicolumn{3}{c}{\textit{Dependent variable:}} \\ 
\cline{2-4} 
\\[-1.8ex] & Log \# Russian apps & Log \# apps & Log \# non-Russians apps \\ 
\\[-1.8ex] & (1) & (2) & (3)\\ 
\hline \\[-1.8ex] 
 $\log p_{t(j)} \times \textsc{RcsInt} =(1.8e-06,0.1]$ & $-$0.002 & $-$0.021 & $-$0.005 \\ 
  & (0.171) & (0.681) & (0.666) \\ 
  & & & \\ 
 $\log p_{t(j)} \times \textsc{RcsInt} =(0.1,0.25]$ & 0.001 & $-$0.010 & 0.003 \\ 
  & (0.171) & (0.595) & (0.584) \\ 
  & & & \\ 
 $\log p_{t(j)} \times \textsc{RcsInt} =(0.25,0.5]$ & 0.004 & 0.015 & 0.028 \\ 
  & (0.168) & (0.563) & (0.552) \\ 
  & & & \\ 
 $\log p_{t(j)} \times \textsc{RcsInt} =(0.5,0.75]$ & 0.102 & $-$0.113 & $-$0.097 \\ 
  & (0.162) & (0.561) & (0.550) \\ 
  & & & \\ 
 $\log p_{t(j)} \times \textsc{RcsInt} =(0.75,1]$ & 0.709$^{***}$ & 0.376 & 0.307 \\ 
  & (0.173) & (0.502) & (0.482) \\ 
  & & & \\ 
\hline \\[-1.8ex] 
Day FE & Y & Y & Y \\ 
Mean outcome (in levels) & 0.3 & 31.81 & 31.51 \\ 
Observations & 303,104 & 303,104 & 303,104 \\ 
R$^{2}$ & 0.216 & 0.007 & 0.008 \\ 
Adjusted R$^{2}$ & 0.216 & 0.007 & 0.008 \\ 
\hline 
\hline \\[-1.8ex] 
\end{tabular}
\\
\begin{minipage}{0.95 \textwidth}
{\footnotesize \emph{Notes}:
This table reports regressions on vacancy level outcomes on changes in the price of the USD (measured in rubles) and its interaction with indicators how Russian-compatible that opening is, based on a model fit with historical data. 
\starlanguage}
\end{minipage}
\end{table}

\begin{figure}
\caption{Elasticities of per-application measures of supply with respect to the price of the US dollar, in rubles}
\label{fig:job_opening_effects}
\centering
\begin{minipage}{1.0 \linewidth}
  \includegraphics[width = \linewidth]{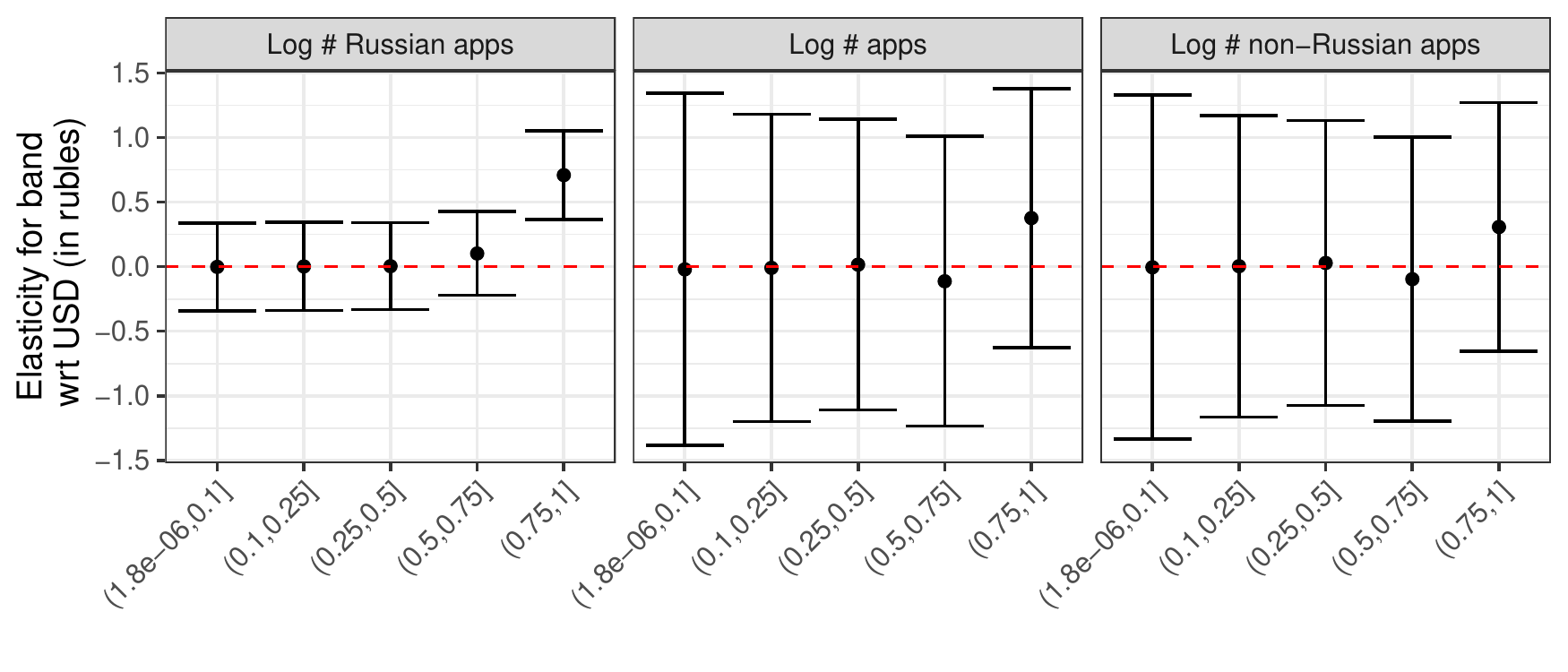} \\
  \begin{footnotesize}
    \begin{singlespace}
      \emph{Notes:} This figure plots the $\beta_k$ coefficients from an estimate of Equation~\ref{eq:job_opening_effects} for several outcomes. 
    \end{singlespace}
    \end{footnotesize}
\end{minipage}
\end{figure}

How this Russian influx translated into the size of the actual applicant pool is considered in Column~(2), which reports the results of a regression where the outcome is the log number of applicants.
The results are highly imprecise and none of the effects are statistically significant.
Similarly, in Column~(3), the outcome is the log number of non-Russian applicants, plus 1 and again, the point estimates are imprecise. 
In short, there is little that can we said about the effects of Russian entrants on applicant pool sizes. 
As it will be more informative to view the collection of elasticities for other outcomes graphically, in Figure~\ref{fig:job_opening_effects}, we re-report the coefficients from Table~\ref{tab:job_opening_apps}.

\begin{figure}
\caption{Elasticities of per-application measures of wage bids and hiring with respect to the price of the US dollar, in rubles}
\label{fig:job_opening_effects_wb_hires}
\centering
\begin{minipage}{1.0 \linewidth}
  \includegraphics[width = \linewidth]{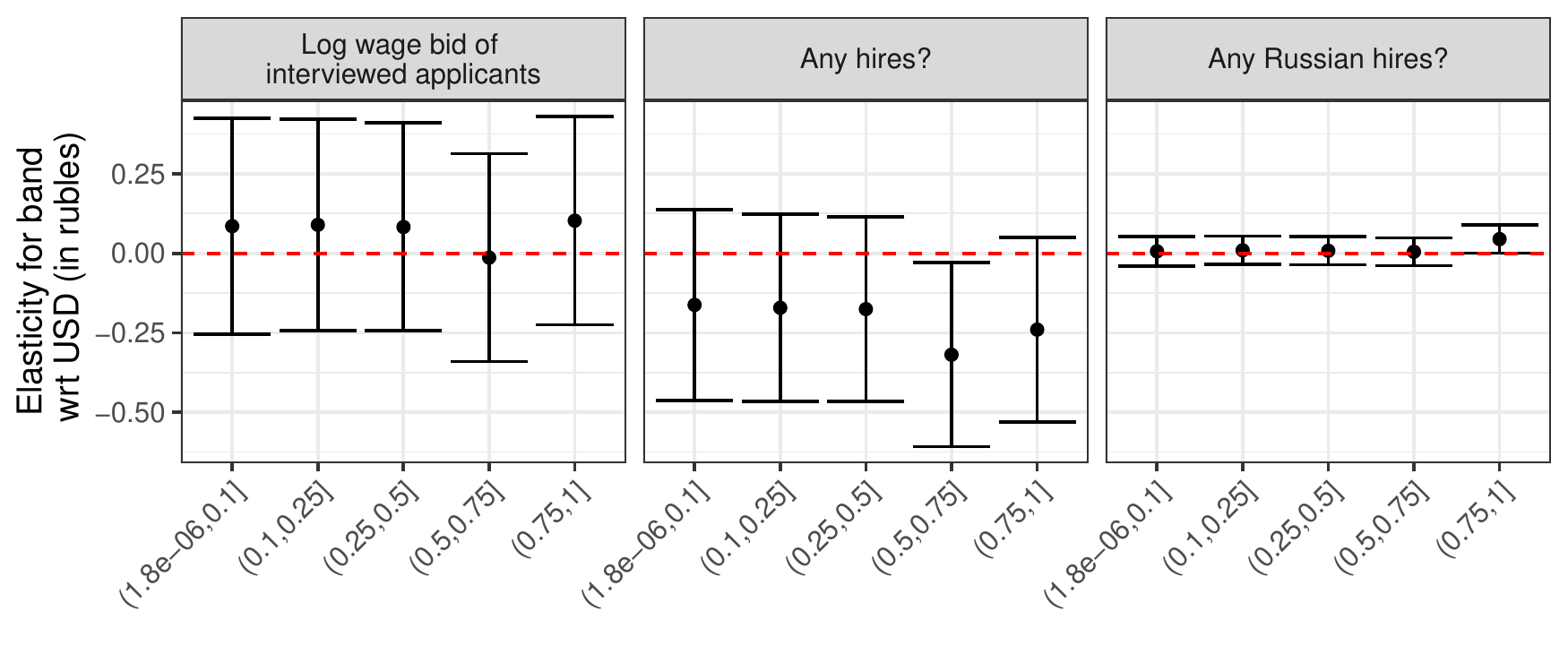} \\
  \begin{footnotesize}
    \begin{singlespace}
      \emph{Notes:}  This figure plots the $\beta_k$ coefficients from an estimate of Equation~\ref{eq:job_opening_effects} for several outcomes. 
    \end{singlespace}
    \end{footnotesize}
\end{minipage}
\end{figure}

Now we turn to wage bidding and hiring.
In Figure~\ref{fig:job_opening_effects_wb_hires}, I turn to the effects of the ruble collapse on measures of wage bids and hiring at the job opening level.
In the leftmost panel, the outcome is the average log wage of interviewed candidates.
There is no evidence of that the ruble collapse had any effect on the wage bid of interviewed applicants, even among the most Russian-compatible openings. 
In the next facet, we look at whether the employer hired anyone at all.
The point estimates are all centered around zero and there is no evidence that more Russian-compatible openings had higher hiring rates. 

In the rightmost panel, the outcome is whether the employer hired a Russian.
Here we see positive effects from about the \nth{75} percentile onward, and large effects at the highest percentile. 

In short, at the level of the applicant, even for job openings most exposed to the Russian influx, the effects on realized wage bids and hiring was non-existent.
The only finding that seems fairly clear is that more Russians were ultimately hired among the most Russian compatible job openings.

\section{Discussion}  \label{sec:discussion}
\begin{figure}
\caption{Simple supply and demand treatment of a Russian supply shock to a global online labor market} 
\label{fig:diagram}
\centering
\begin{minipage}{1.0 \linewidth}
  \includegraphics[width = \linewidth]{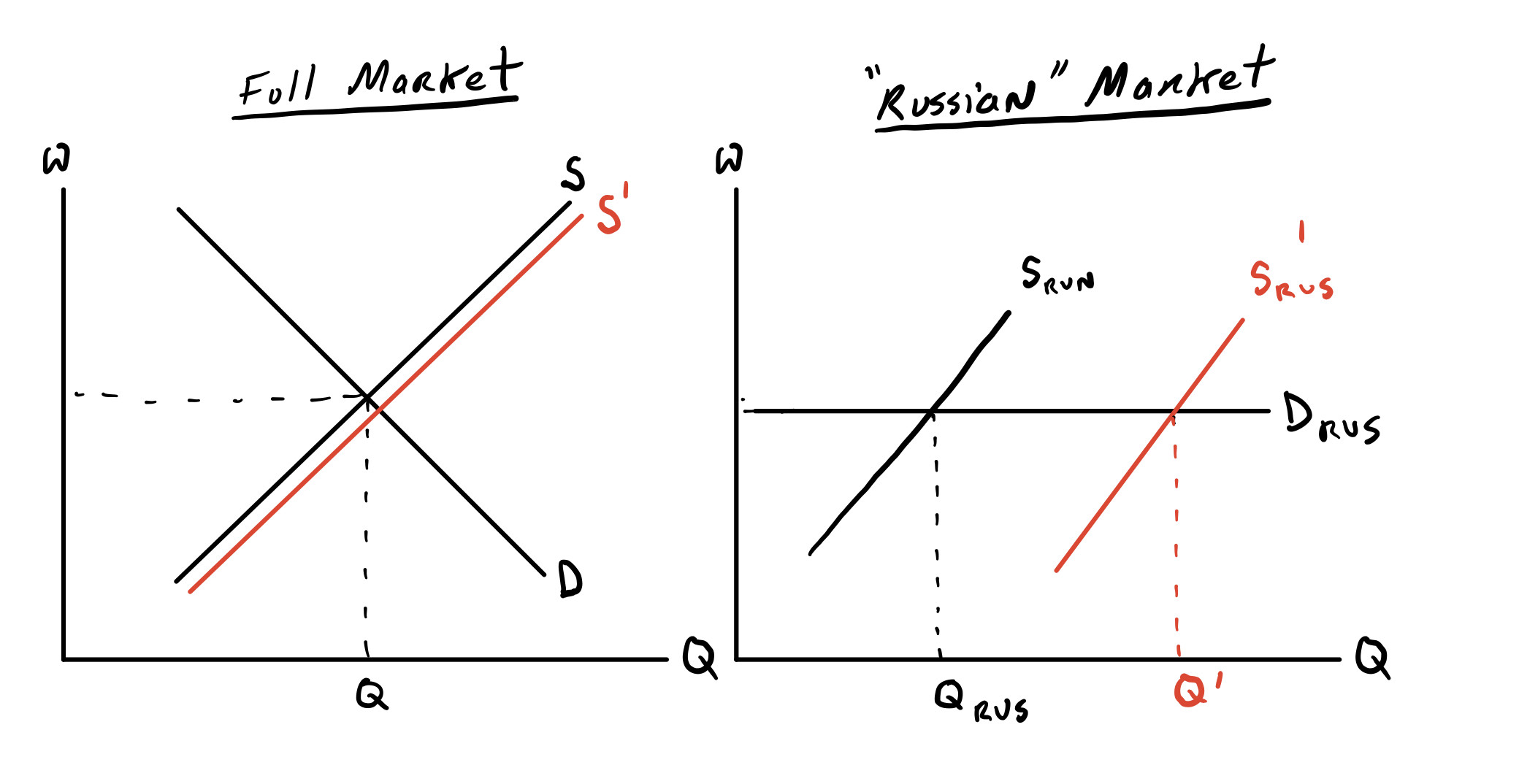} \\
  \begin{footnotesize}
    \begin{singlespace}
    \end{singlespace}
    \end{footnotesize}
\end{minipage}
\end{figure}

One way to characterize the effects of the ruble collapse is with a simple supply and demand model.
In the left panel of Figure~\ref{fig:diagram}, I draw overall supply and demand curves in the market for labor to the platform.
The ruble collapse pushes out the supply curve from $S$ to $S'$, but given that Russians make up a small fraction of the market, it is a small shift.
This shift causes little change to overall quantities or wages.

Now imagine there is also a market for Russian labor specifically, illustrated in the right panel of Figure~\ref{fig:diagram}.
There is a demand curve for Russian labor, but because a would-be buyer of Russian labor has many good substitutes---namely workers from other countries with similar skills---the demand curve is \emph{de facto} horizontal.
In the market for Russian labor, the supply shift from the ruble collapse is obviously much larger.
However, because of the horizontal demand curve, the supply shock leads to a change in quantities ($Q_{RUS}$ to $Q'$) rather than prices. 
This explains the Figure~\ref{fig:country_panel_results} result of large increases in hires (and thus hours-worked) for Russians, but no apparent changes in wages. 
This simple supply and demand explanation succinctly characterizes all of the main results of the paper.

\subsection{What if the supply curve to a marketplace is \emph{de facto} horizontal?} 
The left panel of Figure~\ref{fig:diagram} still draws the market supply curve as upward sloping.
The explanation for why there is no evidence of a Russian decline in wages is that the implied supply shock was small.
But what if this curve was horizontal, in which case even large supply shocks would do little to prices?

Horizontal supply curves in the long-run are found in industries where there are 1) limited capital requirements, 2) more or less free entry and 3) there are constant returns to scale in production.
This seems like a good description of many online marketplaces.
For (1), in most online marketplaces, capital requirements (both physical and human) are minimal or at least common and not specialized (e.g., ride-sharing requires a car).
For (2) platforms try to make it easy to join and remove entry barriers.
And for (3) when the supply side is a large number of smaller sellers, constant returns to scale is reasonable---doubling the number of sellers likely doubles platform output.
Furthermore, even if the industry supply curve was not horizontal, platform competition would tend to make it \emph{de facto} horizontal as market participants multi-home or switch platforms. 

If horizontal supply curves are a reasonable characterization in many online marketplaces, we should expect prices in the market to be quite stable, regardless of demand.
There is some evidence for this from multiple online marketplace contexts.
\cite{hall2017labor} shows Uber driver hourly earnings rates being largely insensitive to the platform imposed price due to a supply response by drivers.
\cite{hortonTambe2020} finds no evidence that workers specializing in Flash saw their wages decrease as the demand for Flash plummeted, as workers quickly switched to other skills.
\cite{cullen20} finds evidence from TaskRabbit---a marketplace for in-person services---that despite huge changes in demand, prices stayed roughly constant and that large changes in scale had little effect on matching efficiency.
\cite{agrawal2015digitization} presents roughly contemporaneous times series evidence on average platform wages, showing remarkable stability despite the platform growing rapidly during this period.

One could imagine the lack of prices effects from these difference cases is the result of a sloped supply curve moving in at exactly the right rate to keep prices constant despite enormous growth---and surely some demand growth has to be met by supply growth---but this seems less plausible than an alternative, which is that supply curve was \emph{de facto} horizontal and so prices never changed despite rapid changes in demand.   

If supply to a platform is highly elastic, it means that sellers will always get about some market-determined rate of return.
If the platform say, pays bonuses of some kind, then under free entry, these would likely get dissipated away by competition. 
This, in a nutshell, is the main finding of \cite{hall2017labor}.
From the profit-motivated market designer's perspective then, the main supply strategy in an established marketplace is to try to lower on-platform costs for sellers.
For example, they could lower costs by creating feature that ease matching frictions or, in the spirit of the two-sided markets literature, try to bring more buyers to the platform.
These benefits would, of course, still be dissipated by competition but can be paired with prices increases or with greater revenue from an expanded market with greater quantities.
In short, the platform strategy in a world of horizontal supply curves is to try to lower costs for buyers and sellers and then raising prices, leaving buyers and sellers with unchanged pay-offs but with the platform's revenue increased.  

\subsection{Implications of the results for the take rate} 

Although the much of the two-sided platform pricing literature has focused on membership fees, many marketplace businesses use ad valorem charges rather than membership/entrance fees.
Despite the theoretical advantages of taxing membership and pricing transactions at cost \citep{oi1971disneyland}---and their use in some settings---many online platforms use ad valorem charges, as they scale with user value (e.g., a person doing 2 rides on Uber would find any fixed membership fee too high if the fee was priced to extract value from the 15 rides per week person).
Furthermore, as the platform intermediates the transaction, it observes and can tax that transaction, which might not be the case in other offline settings (e.g., \cite{jin2015}).
The other advantage of only taxing transactions is that users do not have to know their valuation ex ante.

The ruble episode strongly suggests that lowering the platform ad valorem fees would attract more supply.
However, the economics of this being profit-maximizing are challenging.
Consider that the elasticity of Russians appears to be about 1---a 10\% increase in returns increases hours-worked from Russians by 10\%.
As wages did not change, platform revenue is proportional to hours-worked, as the platform taxes the wage bill (hours-worked times the wage).

If the platform charges say 10\%, even if the increase in hours-worked was entirely incremental (not cannibalizing any other hours from non-Russians), and the platform lowers its fee from 10\% to 9\%, this is a 10\% reduction in revenue but only leads to a 1\% increase in platform earnings and a 1\% increase in hours-worked.
The change in revenue is percentage change in revenue is $0.90 * 1.01 \approx 0.91$.
This of course suggests higher fees would increase revenue.
However, the long-run elasticity is not known and is surely greater than 1, as long-run elasticities are typically larger than short-run elasticities. 
It is not obviously the case than the platform can profitably raise prices, or lower prices for that matter.

Pricing is hard and many platforms choose something simple and tend to stick with it.
While we should not assume what is is necessarily optimal, that so many platform marketplaces choose a simple price structure suggests most platform competition is about features and marketing rather than finely tuned price structures and levels.
This perspective might be quite sensible, particularly for emerging platforms where even small differences in rates of platform growth are likely to be far more important to ultimate platform value. 

\section{Conclusion} \label{sec:conclusion}
For a would-be market designer, the ruble experience offers several lessons.
First, it is clear that workers responded strongly to the financial opportunities created on the platform.
While platforms can and do make ``horizontal'' changes to their platforms to attract buyers and sellers, the ruble experience illustrates the paramount role played by financial returns in explaining market participation---even if lowering the take rate is not a viable way to increase supply, given the relevant elasticities.

Second, the price effects of adding more supply could be safely ignored.
Although the influx of Russians was small relative to the total collection of workers---and so we would not expect market wages to fall---it was conceivable that incoming Russians would have underbid incumbents to get more work, as in bargaining models, potentially lowering platform revenue.
This was not the case.
There was not apparent downside to adding more supply with respect to price.
However, neither was there was there evidence of any positive cross-side externality either---there was no evidence of more Russian friendly job openings being posted.
Given the lack of price effects, this is perhaps unsurprising---though the lack of price effects was less predictable.

In terms of generalization, it is important to note that the ruble-induced influx was not a large supply shock at the market level.
Larger shocks could have different effects.
Furthermore, this was a shock that occurred in an established marketplace well past the chicken-and-egg stage.
Marginal supply for smaller markets could be worth a great deal more. 

\newpage

\bibliographystyle{aer}
\bibliography{russia_shock.bib}

\newpage

\appendix

\newpage

\renewcommand\thefigure{A\arabic{figure}}    
\setcounter{figure}{0}  

\setcounter{table}{0}
\renewcommand{\thetable}{A\arabic{table}}

\section{Online Appendix} \label{sec:online_appendix}

\newpage

\subsection{Table versions} \label{sec:table_versions}

Table~\ref{tab:country_panel_results} has country panel results. 

\begin{table}[!htbp] \centering 
  \caption{Effects of the ruble collapse whether the worker worked any hours in a week} 
  \label{tab:country_panel_results} 
\small 
\begin{tabular}{@{\extracolsep{5pt}}lcccc} 
\\[-1.8ex]\hline 
\hline \\[-1.8ex] 
 & \multicolumn{4}{c}{\textit{Dependent variable:}} \\ 
\cline{2-5} 
\\[-1.8ex] & log(num.active) & log(num.apps) & log(num.hires) & mean.wage \\ 
\\[-1.8ex] & (1) & (2) & (3) & (4)\\ 
\hline \\[-1.8ex] 
 $\log p_t \times \textsc{Russian}_c$ & 1.321$^{***}$ & 1.326$^{***}$ & 1.165$^{***}$ & $-$0.208$^{**}$ \\ 
  & (0.157) & (0.157) & (0.146) & (0.080) \\ 
  & & & & \\ 
 $\log p_t \times RC$ & 1.952 & 3.601 & $-$6.444 & $-$5.232 \\ 
  & (7.787) & (7.246) & (6.871) & (5.029) \\ 
  & & & & \\ 
 $\textsc{Post} \times \underline{w}_c$ & $-$0.005 & 0.051 & 0.100$^{**}$ & $-$0.119$^{***}$ \\ 
  & (0.035) & (0.035) & (0.042) & (0.023) \\ 
  & & & & \\ 
\hline \\[-1.8ex] 
Country FE & Y & Y & Y & Y \\ 
Week FE & Y & Y & Y & Y \\ 
Observations & 549 & 549 & 549 & 549 \\ 
R$^{2}$ & 0.998 & 0.997 & 0.994 & 0.997 \\ 
Adjusted R$^{2}$ & 0.998 & 0.997 & 0.993 & 0.996 \\ 
\hline 
\hline \\[-1.8ex] 
\end{tabular}
\\
\begin{minipage}{1.0 \textwidth}
{\footnotesize \emph{Notes}: 
\starlanguage}
\end{minipage}
\end{table}

Table~\ref{tab:individual_panel_results} has individual panel results. 

\begin{table}[!htbp] \centering 
  \caption{Effects of the ruble collapse on individual outcomes} 
  \label{tab:individual_panel_results} 
\small 
\begin{tabular}{@{\extracolsep{5pt}}lcccc} 
\\[-1.8ex]\hline 
\hline \\[-1.8ex] 
 & \multicolumn{4}{c}{\textit{Dependent variable:}} \\ 
\cline{2-5} 
\\[-1.8ex] & Apps & Hours & Earnings & Avg. Wage \\ 
\\[-1.8ex] & (1) & (2) & (3) & (4)\\ 
\hline \\[-1.8ex] 
 $\log p_t \times \textsc{Russian}_i$ & 0.213$^{***}$ & 0.174$^{***}$ & 0.138$^{**}$ & $-$0.029 \\ 
  & (0.043) & (0.051) & (0.053) & (0.015) \\ 
  & & & & \\ 
\hline \\[-1.8ex] 
Worker FE & Y & Y & Y & Y \\ 
Observations & 251,999 & 213,966 & 211,715 & 211,715 \\ 
R$^{2}$ & 0.435 & 0.611 & 0.681 & 0.940 \\ 
Adjusted R$^{2}$ & 0.363 & 0.592 & 0.666 & 0.937 \\ 
\hline 
\hline \\[-1.8ex] 
\end{tabular}
\\
\begin{minipage}{1.0 \textwidth}
{\footnotesize \emph{Notes}: 
\starlanguage}
\end{minipage}
\end{table}

\newpage

\newpage
\end{document}